\documentstyle[aps,prb,multicol,epsf,amsfonts,amssymb]{revtex}

\newcommand{\mibs}[1]{\mbox{\scriptsize{\bf #1}}}

\newcommand{\citePRSA}[3]{Proc.\ Roy.\ Soc.\ A {\bf #1}, #3 (#2)}
\newcommand{\citePR}[3]{Phys.\ Rev. {\bf #1}, #3 (#2)}
\newcommand{\citePRB}[3]{Phys.\ Rev.\ B {\bf #1}, #3 (#2)}
\newcommand{\citePRL}[3]{Phys.\ Rev.\ Lett. {\bf #1}, #3 (#2)}
\newcommand{\citeRMP}[3]{Rev.\ Mod.\ Phys. {\bf #1}, #3 (#2)}
\newcommand{\citeJPSJ}[3]{J.\ Phys.\ Soc.\ Jpn. {\bf #1}, #3 (#2)}
\newcommand{\citePTP}[3]{Prog.\ Theor.\ Phys. {\bf #1}, #3 (#2)}

\newcommand{\citeEPJB}[3]{Eur.\ Phys.\ J. B {\bf #1}, #3 (#2)}
\newcommand{\citeJPF}[4]{J.\ Phys.\ (France) #1 {\bf #2}, #4 (#3)}
\newcommand{\citeIBID}[3]{ibid. {\bf #1}, #3 (#2)}

\newcommand{\veps}{\varepsilon}

\newcommand{\ave}[1]{{\langle#1\rangle}}

\newcommand{\up}{\uparrow}
\newcommand{\down}{\downarrow}

\begin{document}
\draft
\title{Mott Transitions in the 2D Half-Filled Hubbard Model: Correlator Projection Method with Projective Dynamical Mean-Field Approximation}

\author{Shigeki Onoda$^1$
 and Masatoshi Imada$^2$
}
\address{$^1$Tokura Spin Superstructure Project, ERATO, Japan Science and Technology Corporation, Department of Applied Physics, University of Tokyo, Hongo 7-3-1, Tokyo 113-8656, Japan\\
$^2$Institute for Solid State Physics, University of Tokyo, Kashiwanoha 5-1-5, Kashiwa, Chiba 277-8581, Japan}
\date{recieved}
\maketitle
\noindent
\begin{abstract}
\noindent
\hspace*{-4mm}
The 2D half-filled Hubbard model is studied by a nonperturbative analytic theory of correlator projection. The dynamical mean-field approximation (DMFA) is reproduced at the first-order projection and then improved by systematic inclusion of spatial correlations at higher orders. A geometrical frustration induces a first-order Mott transition surface with a finite-temperature critical end curve and related crossovers. Growth of antiferromagnetic correlations gives single-particle spectra strongly modified from DMFA with shadow bands and flat dispersions observed in high-$T_{\rm c}$ cuprates.
\end{abstract}


\begin{multicols}{2}
\narrowtext
\sloppy

Analytic description of metal-insulator transitions (MIT) driven by electron correlation postulated by Mott~\cite{Mott} has been one of the most intriguing issues in condensed-matter physics. This is also required for understanding high-$T_{\rm c}$ superconductivity~\cite{BednortzMuller86} that emerges around the MIT. In spite of substantial progress in the long history of studies~\cite{RMP_Imada,Hubbard,BrinkmanRice70,Slater,DMFT}, complete theoretical descriptions of electronic states around the MIT remain open.

The dynamical mean-field (DMF) theory~\cite{DMFT} unifies the Mott-Hubbard~\cite{Hubbard} and Brinkman-Rice~\cite{BrinkmanRice70} pictures within the Fermi liquid. In infinite dimensions where the theory becomes exact~\cite{DMFT}, the first-order MIT line with a finite-temperature critical end point~\cite{Cyrot72,Castellani79} has been obtained for the Hubbard model on the fully frustrated Bethe lattice~\cite{RozenbergChitraKotliar99,DMFT}. In realistic dimensions, however, momentum dependence of the self-energy part and spin correlations ignored in the DMF approximation (DMFA) play other important roles near the MIT. Several studies have been done to restore such nonlocal correlations~\cite{OPM,DCA,CDMF}.

Much progress has been made on understandings of filling-control MIT in low dimensions by the operator projection method (OPM)~\cite{OPM}. Based on the unique series of Dyson equations or continued-fraction expansion~\cite{OP} of the Green's function, the method systematically improves the mean-field, conserving~\cite{KadanoffBaym,FLEX,TPSC} and Hubbard-type~\cite{Hubbard,Roth69,Gros94} approximations. By taking account of spin correlations as well as the Hubbard-band splitting, this gives a unified scheme of the filling-control MIT beyond the early pictures~\cite{Mott,Hubbard,BrinkmanRice70,Slater}. This theory also reproduces numerically inferred key elements~\cite{FurukawaImada,BulutScalapinoWhite94,HankePreuss95Grober00,Assaad}; the diverging compressibility towards the MIT~\cite{FurukawaImada}, the four-band-like structure~\cite{HankePreuss95Grober00} and flat dispersions around the momentum $(\pi,0)$~\cite{Assaad} observed in high-$T_{\rm c}$ cuprates~\cite{ARPES}. However, in  particle-hole asymmetric cases, the self-consistent decoupling approximation~\cite{OPM} fails to reproduce a Mott insulator~\cite{OnodaImadaUNPUBLISHED} due to insufficient estimates of local dynamics.

In this letter, we propose an alternative method to Refs.~\cite{OPM,DCA,CDMF} to take account of momentum dependences ignored in the DMFA. Combining the OPM with a generalized DMFA, we for the first time obtain an MIT phase diagram of the 2D half-filled Hubbard model in the parameter space of the local Coulomb repulsion $U$, the second-neighbor transfer $t'$ and the temperature $T$ scaled by the nearest-neighbor transfer $t$. $t'$ introduces a geometrical frustration and a particle-hole asymmetry. Our results provide a comprehensive picture of the Mott transition including the single-particle dynamics observed in insulating cuprates~\cite{Ino00}. 

At the first-order projection, our formalism hereafter called the correlator projection method (CPM) is reduced to the original DMF theory~\cite{DMFT}. With higher-order projections, spatial correlations are analytically restored in a systematic fashion. Here, the limit of the low energy and high momentum resolution is approached by simultaneously including the lower-energy and longer-ranged spatial correlations along the same concept as the renormalization group. This contrasts with the cluster methods~\cite{DCA,CDMF} where the limit of the high momentum resolution is taken after the low-energy limit. 

At the second-order projection, our projective DMFA gives the first-order MIT surface, critical end curve at $T>0$, crossover between Mott-Hubbard and Slater insulators and metal-insulator crossover at higher $T$'s in the $(U,t',T)$ space. Furthermore, beyond the original DMFA~\cite{DMFA}, antiferromagnetic (AF) or spin-singlet correlations produce the AF shadow structure and flat dispersions observed in high-$T_{\rm c}$ cuprates~\cite{ARPES,Ino00}.

This framework essentially satisfies the Fermi-liquid properties with a large Fermi surface and vanishing damping rates toward the Fermi energy and momenta for metal without any symmetry breaking, in contrast to previous equation-of-motion approaches~\cite{Hubbard,Roth69,Gros94}.
In contrast to cluster methods~\cite{DCA,CDMF}, the CPM is free from cluster-size effects. It also gives an exact uniform solution of the soluble model with electron tranfer and exchange coupling between the nearest neighbors for $SU(N)$ spins with $N\to\infty$~\cite{AffleckMarston88} in 1D, while cluster methods give only approximate solutions~\cite{BiroliKotliar01}.

We consider the Hubbard Hamiltonian $H\equiv-t\sum_{\mibs{x},\mibs{x}',s}^{{\rm n.n.}}c^\dagger_{\mibs{x}s}c_{\mibs{x}'s}-t'\sum_{\mibs{x},\mibs{x}',s}^{{\rm n.n.n.}}c^\dagger_{\mibs{x}s}c_{\mibs{x}'s}-\mu\sum_{\mibs{x}s}n_{\mibs{x}s}+U\sum_{\mibs{x}}n_{\mibs{x}\up}n_{\mibs{x}\down}$ on a square lattice, with the electron creation, annihilation and number operators at a site $\bf{x}$ with a spin index $s$, $c^\dagger_{\mibs{x}s}$, $c_{\mibs{x}s}$ and $n_{\mibs{x}s}$, respectively. $\sum^{{\rm n.n.}}$ and $\sum^{{\rm n.n.n}}$ denote the summations over the nearest and next-nearest neighbors, respectively. $\mu$ is the chemical potential. In two dimensions, the magnetic phases do not appear at $T>0$. The other symmetry breakings are assumed not to occur.

\begin{figure}[tbh]
\begin{center}\leavevmode
\epsfxsize=8.4cm
$$\epsffile{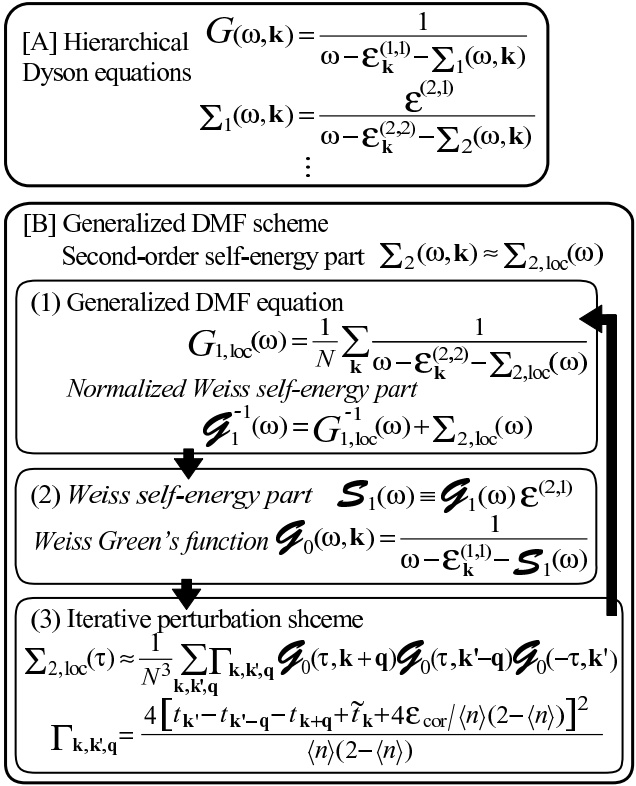}$$
\end{center}
\caption{Schematic self-consistent loop of the currently proposed formalism of CPM. $N$ is the number of the sites. The CPM uses the generalized DMFA to calculate the local dynamics of $\Sigma_n(\omega,{\bf k})$, instead of the self-consistent decoupling approximation in the OPM~\cite{OPM}. The first-order CPM is reduced to a conventional DMFA~\cite{DMFT}.}
\label{fig:formalism}
\end{figure}

The formalism of the CPM is based on the formal and hierarchical Dyson equations noted in the top panel [A] of Fig.~\ref{fig:formalism}. They are obtained by the projection in the equation of motion for $c_{\mibs{x}s}$~\cite{OPM}. 
The explicit definitions for $\veps^{(1,1)}_{\mibs{k}}$, $\veps^{(2,1)}$ and $\veps^{(2,2)}_{\mibs{k}}$ have already been given from the Hubbard parameters $t$, $t'$ and $U$ and equal-time correlations~\cite{OPM,Roth69}. Here, $\veps^{(1,1)}_{\mibs{k}}$ simply yields the Hartree-Fock dispersions.

To solve this continued-fraction expansion form requires an approximation to the highest-order self-energy part $\Sigma_n(\omega,{\bf k})$. Here, we restrict ourselves up to the second-order projection where $n=2$. In earlier studies, $\Sigma_2$ has been calculated by a two-site method~\cite{MatsumotoMancini97} and a self-consistent decoupling approximation~\cite{OPM}. However, with these methods, it is difficult to discuss the MIT when a particle-hole asymmetry exists at half filling because of insufficient treatment of the local dynamics. Instead, we employ a generalized DMFA to correctly calculate the local dynamics of $\Sigma_2(\omega,{\bf k})$ by ignoring its ${\bf k}$ dependence.  This is a systematic extension of the conventional DMFA since $\Sigma_1(\omega,{\bf k})$ contains ${\bf k}$ dependences.

The generalized DMF scheme is illustrated in the flow lines in the bottom panel [B] of Fig.~\ref{fig:formalism}: First, from an arbitrary $\Sigma_{2,{\rm loc}}$, the local {\it normalized} self-energy part $G_{1,{\rm loc}}(\omega)$ is calculated in the DMF procedure (1). Then, the {\it normalized Weiss self-energy part} ${\cal G}_1$ is obtained from $G_{1,{\rm loc}}$ and $\Sigma_{2,{\rm loc}}$. As in the procedure (2), the {\it Weiss self-energy part} ${\cal S}_1$ directly obtained from ${\cal G}_1$ generates the {\it Weiss Green's function} ${\cal G}_0$ by replacing $\Sigma_1$ with ${\cal S}_1$ in the Dyson equation. To calculate $\Sigma_{2,{\rm loc}}(\omega)$, we employ the iterative perturbation scheme~\cite{DMFT} using ${\cal G}_0(\omega,{\bf k})$ in the procedure (3). Then, the loop continues to (1) by replacing the original $\Sigma_{2,{\rm loc}}$ with $\Sigma_{2,{\rm loc}}$ obtained in the procedure (3) until the convergence of $\Sigma_{2,{\rm loc}}$ is reached. A self-consistent solution is obtained after the iteration of this loop. The converged $\Sigma_2(\omega)$ is used to calculate $G(\omega,{\bf k})$ by substituting into the [A]. We also note that in this loop, conservation laws for $G(\omega,{\bf k})$ and $G(\omega,{\bf k})\Sigma_1(\omega,{\bf k})$ are satisfied.

In the loop, $4\veps_{{\rm cor}}/\langle n\rangle(2-\langle n\rangle)$ is defined as the local part of $\veps^{(2,2)}_{\mibs{k}}$. This together with $\veps^{(2,1)}$ gives a source of the Hubbard band splitting. $-t_{\mibs{k}}$ and $-\tilde{t}_{\mibs{k}}$ are the nonlocal parts of $\veps^{(1,1)}_{\mibs{k}}$ and $\veps^{(2,2)}_{\mibs{k}}$, respectively. $\tilde{t}_{\mibs{k}}$ introduces a crucial ${\bf k}$ dependence of $\Sigma_1(\omega,{\bf k})$ mainly through the superexchange interaction. The equal-time charge, spin and local-pair susceptibilities appearing in $\tilde{t}_{\mibs{k}}$ may be determined self-consistently with the solutions of coupled correlator projection equations for {\it two-particle} operators. Here, however, we independently determine them from the two-particle self-consistent method~\cite{TPSC}.

\begin{figure}[tbh]
\begin{center}\leavevmode
\epsfxsize=6.5cm
$$\epsffile{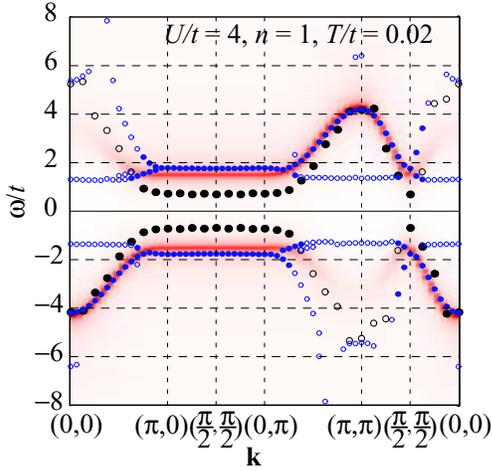}$$
\end{center}
\caption{(color) A density plot of the present results for $A(\omega,{\bf k})$ is shown in red color. Blue and black circles denote the previous results of the OPM with a self-consistent decoupling approximation~\cite{OPM} and QMC results~\cite{Assaad}, respectively. The filled and the open symbols represent the momenta where the peak intensity is more and less than $10{\%}$ of the maximum intensity, respectively.}
\label{fig:Akw2eL1.t0U4T.02}
\end{figure}

Now we show results from the above second-order CPM with projective DMFA. They reproduce important Mott-insulating character of single-particle spectra $A(\omega, {\bf k})\equiv -\frac{1}{\pi}{\rm Im}\ G(\omega, {\bf k})$, as shown in Fig.~\ref{fig:Akw2eL1.t0U4T.02} for $t'=0$: It produces a Mott gap $\Delta$ separating the two Hubbard bands where $\Delta$ seems to grow from 0 with increasing $U$. At low $T$'s, AF or singlet correlations yield AF shadow structure and flat dispersions around the $(\pi,0)$ and $(0,\pi)$ momenta~\cite{OPM,Assaad,ARPES,Ino00} in the Hubbard bands. Such significant modifications at the second-order CPM from the conventional DMFA~\cite{DMFA} are due to inclusion of spatial correlations mainly through superexchange interaction. Besides, present results reproduce a direct gap for $t'=0$ at $T/t=0.02$. $A(\omega,{\bf k})$ shows a remarkable similarity to the QMC results~\cite{Assaad}. Deviations of the momentum distribution at $T/t=0.02$ from $T=0$ QMC data~\cite{Assaad} turn out to be less than $10\%$ for $U/t=4$.
%
%

\begin{figure}[tbh]
\begin{center}\leavevmode
\epsfxsize=6.5cm
$$\epsffile{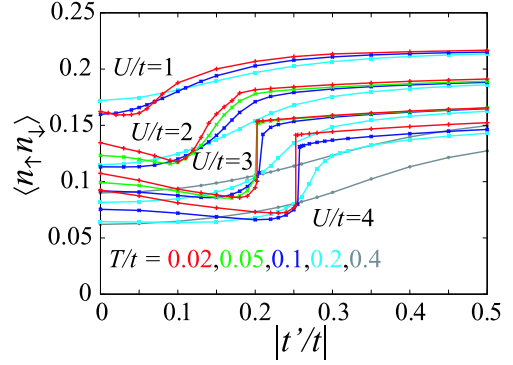}$$
\end{center}
\caption{(color) $\langle n_\uparrow n_\downarrow\rangle$ versus $|t'/t|$ at several $U$'s and $T$'s.}
\label{fig:double}
\end{figure}

\begin{figure}[tbh]
\begin{center}\leavevmode
\epsfxsize=6.0cm\
$$\epsffile{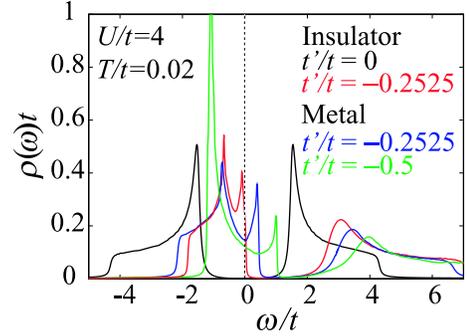}$$
\end{center}
\caption{(color) Density of states $\rho(\omega)$ corresponding to metallic and Mott-insulating solutions.}
\label{fig:DOSeL1.t0_.2525MI_.5U4}
\end{figure}

Next, we discuss the MIT as a function of $t'$ at half filling. The double occupancy $\langle n_\uparrow n_\downarrow\rangle$ exhibits a jump or a singularity depending on the first-order or second-order bandwidth-control MIT from arguments on the MIT in the effective action for spins and charges~\cite{Castellani79,RozenbergChitraKotliar99}.
$\langle n_\uparrow n_\downarrow\rangle$ is plotted against $|t'/t|$ in Fig.~\ref{fig:double}. With decreasing $T$, the maximum slope of $\langle n_\uparrow n_\downarrow\rangle$ versus $|t'/t|$ increases. For $U/t=3$ and $4$, $\langle n_\uparrow n_\downarrow\rangle$ exhibits a jump at $|t'|=t'_{\rm MIT}(U,T)$ below the critical temperature $T_{\rm cr}(U)$. (Alternatively, if we vary $U$ with $t'$ being fixed, $\langle n_\uparrow n_\downarrow\rangle$ exhibits a jump at some value $U_{\rm MIT}(t',T)$.) The jump indicates a first-order transition~\cite{Castellani79}. More correlated phase for $|t'|<t'_{\rm MIT}(U,T)$ is insulating while less correlated phase with larger $\langle n_\uparrow n_\downarrow\rangle$ for $|t'|>t'_{\rm MIT}(U,T)$ is metallic. This is illustrated for the density of states $\rho(\omega)\equiv\frac{1}{N}\sum_{\bf k}A(\omega, {\bf k})$ for $U/t=4$ at $T/t=0.02$ in Fig.~\ref{fig:DOSeL1.t0_.2525MI_.5U4}. Here, an MIT occurs around $t'/t=-0.2525$ with abrupt shifts in the spectral weights from the upper Hubbard band to the lower. The hysteresis seems to be suppressed; the higher-order projection enables us to escape from a metastable solution. Due to the particle-hole asymmetry introduced by $t'$, the Kondo resonance does not appear in a particle-hole symmetric manner as in the fully frustrated Bethe lattice~\cite{DMFT}. This first-order MIT below $T_{\rm cr}$ accompanies a discontinuous change in single-particle dispersions, while the dispersions continuously evolve at $T\ge T_{\rm cr}$. The insulating dispersion with a choice of $t=0.25$ eV agrees with that of La${}_2$CuO${}_4$~\cite{Ino00}.

In this formalism, spatial correlations are restored as the ${\bf k}$ dependence of $\Sigma_1(\omega,{\bf k})$. Neglecting the ${\bf k}$ dependence of $\Sigma_n(\omega,{\bf k})$ with finite $n\ge2$ does not automatically guarantee the Luttinger sum rule except the trivial case of $n=1$: Including the ${\bf k}$ dependence of $\Sigma_1(\omega,{\bf k})$ in the case of $n\ge2$ modifies the shape of the Fermi surface and does not necessarily keep the Luttinger volume, if $\Sigma_1(\omega,{\bf k})$ in the both limits of low energy and high momentum resolutions is not exactly obtained for finite $n$. At $n=1$, the Fermi surface remains the same due to the absence of the $k$ dependence of $\Sigma_1(\omega,{\bf k})$. However, our metallic solutions satisfy the Luttinger sum rule with less than $16\%$ deviations.

As shown in Fig.~\ref{fig:double}, at low temperatures in the insulating region, $\ave{n_\up n_\down}$ increases with decreasing temperature for large $U/t$ and small $|t'/t|$. Furthermore, $\ave{n_\up n_\down}$ decreases with increasing $|t'/t|$. Namely, $\ave{n_\up n_\down}$ seems to increase when the AF correlation is well developed~\cite{PaivaScalettarHuscroft01}. This is interpreted from the following fact: With well developed AF correlations, the coherence of quasiparticles (or in other words, coherent propagation of holons and doublons) increases, which results in the kinetic energy gain thus allows larger double occupancy.

Fig. 5 shows the critical behaviors of the double occupancy; its jump at the first-order MIT ${\mit\Delta}\ave{n_\uparrow n_\downarrow}$ and the inverse of the maximum slope $q=[|d\ave{n_\uparrow n_\downarrow}/d(t'/t)|]_{|t'|=t'_*(U, T)}$ obtained from Fig. 3. For fixed U, a critical point $(t'_{\rm cr}(U), T_{\rm cr}(U))$ exists as a $T>0$ end point of the first-order MIT line $(t'_{\rm MIT}, T_{\rm MIT})$. The least square fits show that $q$ diverges as $(T-T_{\rm cr}(U))^{-1.0}$ for $T>T_{\rm cr}(U)$, while ${\mit\Delta}\ave{n_\uparrow n_\downarrow}$ vanishes as $(T_{\rm cr}(U)-T)^{0.23}$ for $T<T_{\rm cr}(U)$. Similar results have been obtained in infinite dimensions~\cite{RozenbergChitraKotliar99}.

Finally, the MIT phase diagram obtained for the 2D half-filled Hubbard model is shown in Fig.~\ref{fig:phase_tUT} with the first-order MIT surface (blue surface), critical end line (red curve), and metal-insulator crossover (green surface) that is determined from the temperature where the double occupancy varies most rapidly for fixed $U$ and $t'$. $U_{\rm cr}$ and $T_{\rm cr}$ vanish for $t'=0$ and increases with increasing $|t'|$ and/or $T$. Only at moderate $t'$, the MIT phase diagram becomes similar to the DMFA results~\cite{DMFT,RozenbergChitraKotliar99}. The first-order character seems to persist at $T=0$ due to absence of the particle-hole symmetry, unlike the particle-hole symmetric Bethe lattice~\cite{DMFT}. The MIT phase boundary at $T\to0$ agrees with recent numerical results~\cite{KashimaImada01}.
%

In summary, a formalism to systematically include spatial correlations into the dynamical mean-field theory has been given. Mott transitions and the phase diagram of the 2D half-filled Hubbard model have been clarified by this correlator projection method. Local AF or singlet correlations prominently yield AF shadow bands, flat dispersions around $(\pi,0)$ and $(0,\pi)$ and a Mott insulator even at weak couplings, in agreement with the numerical results~\cite{BulutScalapinoWhite94,HankePreuss95Grober00,Assaad}. In the phase diagram in the $(U,|t'|,T)$ space, there exists a first-order MIT surface accompanied by a discontinuous change of single-particle dispersions. The surface ends at a $T>0$ critical curve and a metal-insulator crossover appears at higher $T$'s. The present theory qualitatively reproduces the dispersion in undoped cuprates.

The work was supported by the Japan Society for the Promotion of Science under grant number JSPS-RFTF97P01103.

\begin{figure}[htb]
\begin{center}\leavevmode
\epsfxsize=5.5cm
$$\epsffile{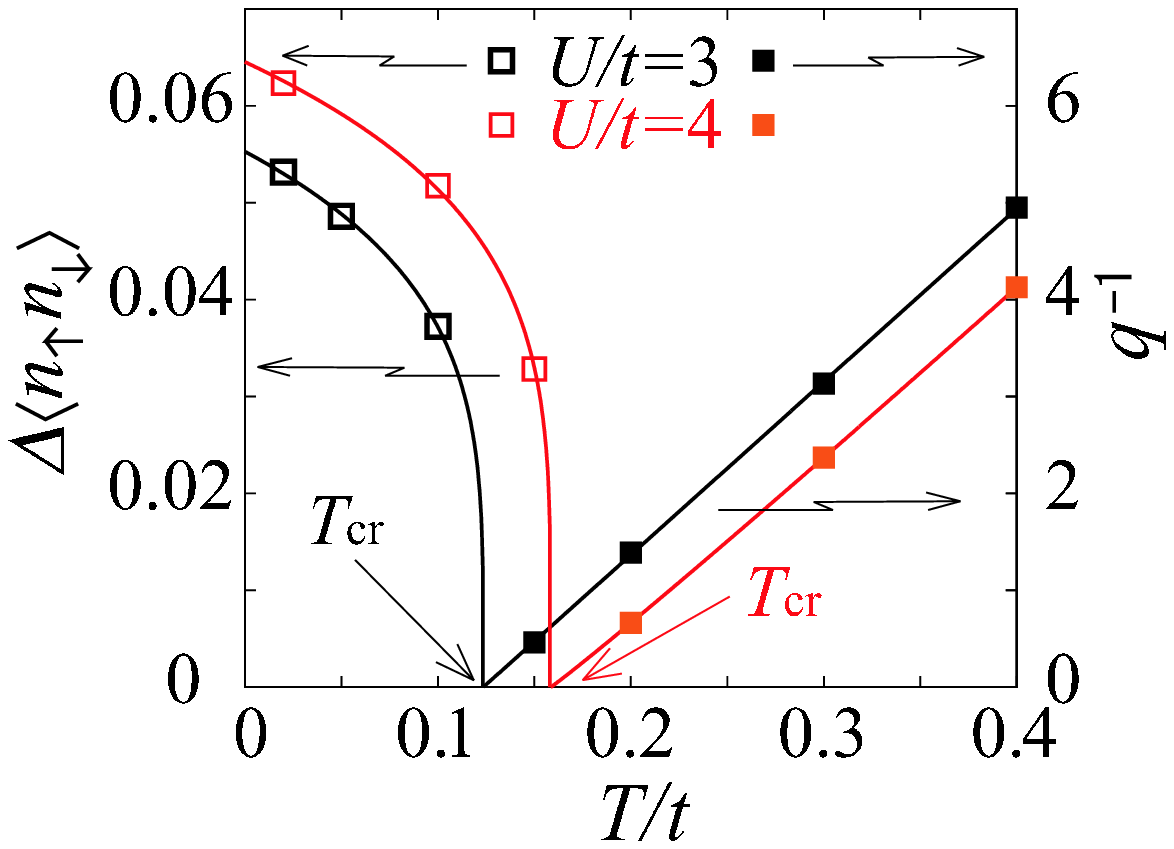}$$
\end{center}
\caption{(color) The jump in $\langle n_\uparrow n_\downarrow\rangle$ (open symbols) and the inverse of its maximum slope $q$ (filled symbols) versus $|t'/t|$ are plotted against $T/t$ with scaling functions. This gives a $T>0$ second-order critical point $T_{\rm cr}$ as an end point of the first-order MIT line $(T_{\rm cr}(U),t'_{\rm cr}(U))$ with $t'_{\rm cr}(U)=t'_*(U,T_{\rm cr}(U))$.}
\label{fig:jump-slope}
\end{figure}
\begin{figure}[htb]
\begin{center}\leavevmode
\epsfxsize=6.0cm
$$\epsffile{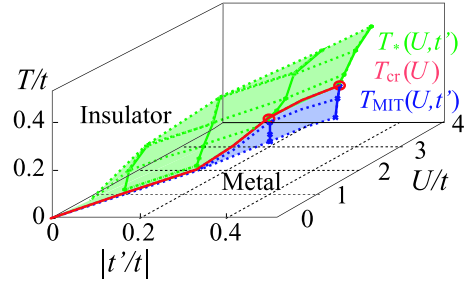}$$
\end{center}
\caption{(color) Phase diagram of the 2D half-filled Hubbard model obtained by the present theory. The critical curve exists as illustrated by the red line. The green surface denoted as $T_*$ gives a metal-insulator crossover.}
\label{fig:phase_tUT}
\end{figure}
\end{multicols}

\end{document}